\documentclass{moriond}

\def\Journal#1#2#3#4{{#1} {\bf #2}, #3 (#4)}

\def\PRD{{\em Phys. Rev.} D}

\def\be{\begin{equation}}
\def\ee{\end{equation}}
\def\bea{\begin{eqnarray}}
\def\eea{\end{eqnarray}}

\newcommand{\Photo}{\includegraphics[height=35mm]{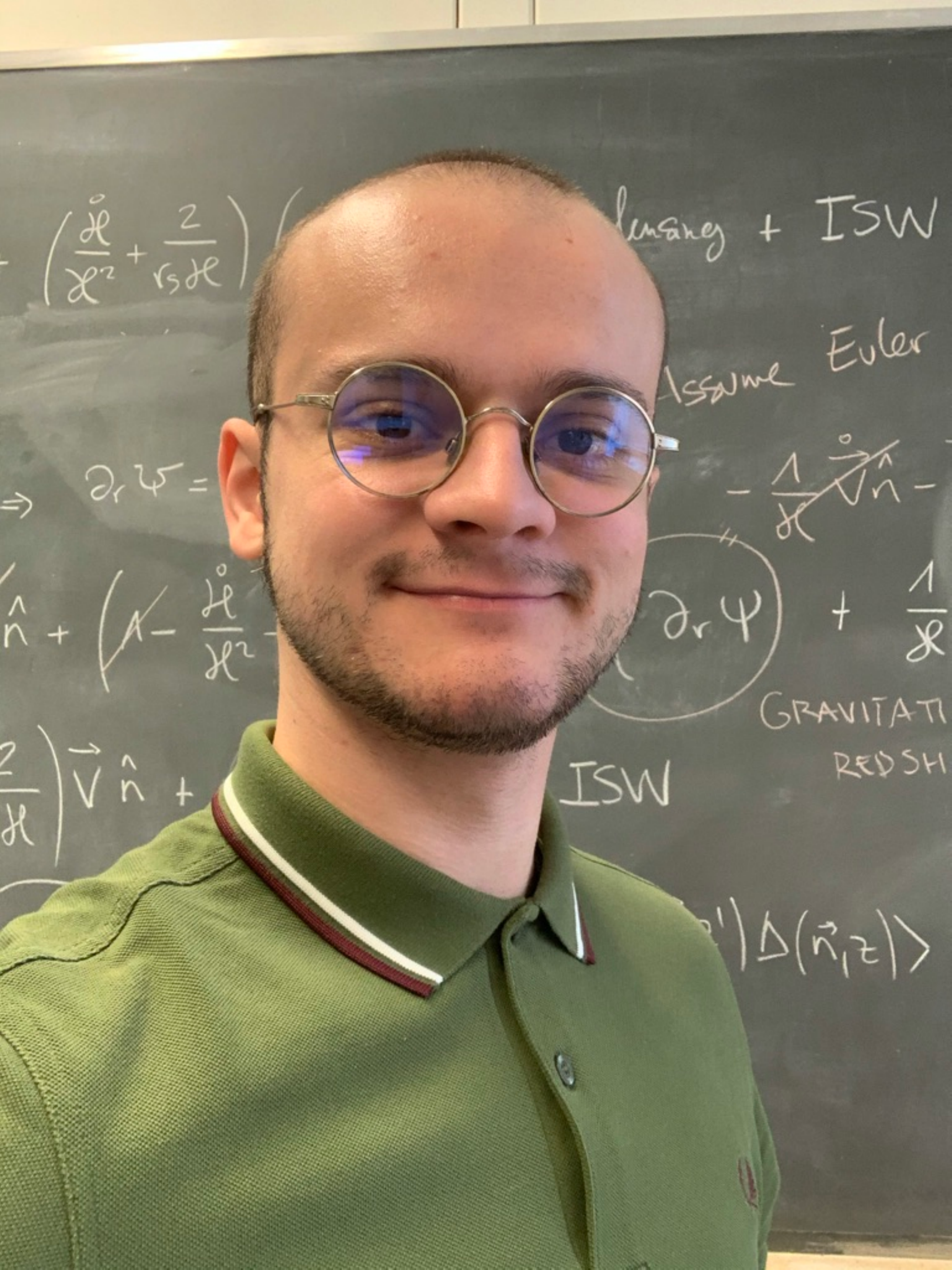}}

\begin{document}
\vspace*{4cm}
\title{ESTIMATOR FOR THE ANISOTROPIC STRESS USING RELATIVISTIC EFFECTS IN LARGE-SCALE STRUCTURE \footnote{Contribution to the 2022 Cosmology session of the ``56th Rencontres de Moriond". }}

\author{ Daniel SOBRAL BLANCO \footnote{E-mail: daniel.sobralblanco@unige.ch}}

\address{University of Geneva, Department of Theoretical Physics,  24, quai Ernest Ansermet \\
1211 Genève 4
Switzerland}

\maketitle\abstracts{
The large-scale structure of the Universe is a rich source of information to test the consistency of General Relativity on cosmological scales. We briefly describe how the observed distributions of galaxies is affected by redshift-space distortions, but also by gravitational lensing and other relativistic effects. Then, we show how one of this relativistic effects, the gravitational redshift, can be used to build a model independent test that directly measures the anisotropic stress, i.e. the difference between the two gravitational potentials that describe spacetime fluctuations of the geometry.} 

\section{Introduction}

Testing the laws of gravity at cosmological scales is one of the main science drivers for the coming generation of large-scale structure surveys. At large scales and late times, our Universe can be consistently described by a set of four fields that quantify the deviations from an homogeneous and isotropic background: the two metric potentials, $\Phi$ and $\Psi$, describing the fluctuations in the geometry of the Universe, the matter density fluctuations, $\delta$, and the galaxy peculiar velocity field, $\mathbf{V}$ \footnote{We use the metric convention $ds^2=a^2(t)[-(1+2\Psi)dt^2+(1-2\Phi)d\mathbf{x}^2]$ where $t$ denotes conformal time.}. Testing the laws of gravity requires then to test the relations between these four fields. This can be done by using the data obtained at galaxy surveys, which probe the late-time large scale structure of the Universe. 

One of this test of gravity consists in measuring the ratio between the two metric potentials, characterized by the \textit{anisotropic stress}, $\Phi=\eta\Psi$. $\Lambda$CDM and many dark energy models predict $\eta=1$ but, generally, modified gravity models predict $\eta\neq 1$. However, the observables at late-times considered so far are sensitive to only three combinations of the four fields, namely $\delta$, $\mathbf{V}$ and $(\Phi+\Psi)$. This means that current observations are not able to test all relations between the four fields. To overcome this problem, one usually has to assume that the continuity equation for dark matter holds: there is no exchange of energy between dark matter and dark energy; and that Euler equation also holds: dark matter follows geodesics. Using this, a measurement of $\mathbf{V}$ can be translated into a measurement of $\Psi$, which can then be compared to $(\Phi+\Psi)$ to test if the two potentials are the same \cite{amen}. We argue that this procedure only provides an indirect test that fails to give a measurement of the anisotropic stress if any of the aforementioned assumptions is not valid \cite{own}. 

In this talk, we give a gauge-independent observable for galaxy clustering at linear order in perturbation theory which include relativistic effects. These terms will become relevant for future generations of galaxy surveys. Given that the relativistic effects are sensitive to  $\Psi$, it is thus possible to isolate and directly measure it in a model independent way. We finally show how we can build a new estimator for the anisotropic stress. 

\section{Galaxy number counts at linear order}

Galaxy surveys probe the large-scale structure of the Universe at late-times. They provide us with maps of the sky in which the position of the galaxies can be parametrized in terms of redshift $z$ and the direction of observation $\mathbf{n}$. We can then \textit{count} the number of galaxies per pixel and measure the fluctuations between different pixels. Our observable is the \textit{galaxy number counts}, defined as $\Delta(z,\mathbf{n})=(N-\bar{N})/{\bar{N}}$, where $N$ is the number of galaxies at the pixel $(z,\mathbf{n})$ and $\bar{N}$ the average number of galaxies per pixel. The question is now: how is $\Delta$ related to the initial perturbations, the theory of gravity and dark energy? In order to answer this question, we calculate the perturbed photon geodesics to infer the change in energy and the change in direction induced by the underlying structure, which induce distortions in the coordinate system of $(z,\mathbf{n})$. The final expression at linear order in perturbation theory reads \cite{bon,yoo,chall}
\begin{eqnarray}
\label{eq:numbercounts}
\Delta(z,\mathbf{n})&=&b\delta-\frac{1}{\mathcal{H}}\partial_r(\mathbf{V}\cdot\mathbf{n})+(5s-2)\int^r_0dr'\frac{r-r'}{2rr'}\Delta_\Omega(\Phi+\Psi) \\
&+& \left(1-5s-\frac{\dot\mathcal{H}}{\mathcal{H}^2}+\frac{5s-2}{r\mathcal{H}}\right)\mathbf{V}\cdot\mathbf{n}+\frac{1}{\mathcal{H}}\dot\mathbf{V}\cdot\mathbf{n}+\frac{1}{\mathcal{H}}\partial_r\Psi \nonumber \\
&+& \frac{2-5s}{r}\int^r_0dr'(\Phi+\Psi)+3\mathcal{H}\nabla^{-2}(\nabla\mathbf{V})+\Psi+(5s-2)\Phi\nonumber \\
&+& \frac{1}{\mathcal{H}}\dot\Phi+\left(\frac{\dot\mathcal{H}}{\mathcal{H}^2}+\frac{2-5s}{r\mathcal{H}}+5s\right)\left[\Psi+\int^r_0dr'(\dot\Phi+\dot\Psi)\right], \nonumber
\end{eqnarray}
Where $b$ is the galaxy bias, $s$ is the magnification bias and $\mathcal{H}=a H$ is the comoving Hubble factor. The two first terms in the first line are the dark matter density perturbations and the \textit{redshift-space distortions} \cite{kai,ham}~\footnote{We call the combination of both as the \textit{standard} terms, since they are the relevant contributions for current measurements of $\mathbf{V}$ and $\delta$.}, while the third term is the \textit{weak-lensing} contribution. The latter is only relevant for very high redshifts and can be neglected for our purposes. The rest of the terms encode the \textit{relativistic distortions}. The second line contain two Doppler effects and the \textit{gravitational redshift} contribution, which is proportional to $\partial_r \Psi$ \footnote{Notice that we are not using Euler equation here. If we use it, the three terms in the second line combine to a single Doppler term. We want to keep the gravitational redshift explicit and the most general form of $\Delta$.}. The latter can be isolated to provide a direct measurement of the metric potential $\Psi$ \cite{own}. The third and fourth lines are other relativistic distortions involving the scalar metric potentials. 

\section{Isolating $\Psi$ in the correlation function}

In practice, we do not measure the fluctuations in the number counts of galaxies per pixel individually. We study the correlation function between pairs of pixels and perform a multipole expansion. Using Eq.(\ref{eq:numbercounts}), the correlation function will be of the form
\begin{eqnarray}
    \xi(z,d)&=&\langle\Delta(\mathbf{x})\Delta(\mathbf{x}')\rangle \nonumber \\
    &=&\langle\Delta_{st}(\mathbf{x})\Delta_{st}(\mathbf{x}')\rangle+\langle\Delta_{st}(\mathbf{x})\Delta_{rel}(\mathbf{x}')\rangle+\langle\Delta_{rel}(\mathbf{x})\Delta_{st}(\mathbf{x}')\rangle
    +\langle\Delta_{rel}(\mathbf{x})\Delta_{rel}(\mathbf{x}')\rangle,
\end{eqnarray}
where we wrote $\mathbf{x}=(z,\mathbf{n})$ for simplicity. The correlation function will only depend on the redshift $z$ and the relative distance between the pixels, $d$. The last term is subdominant and can be neglected for our purposes. The relativistic corrections to the correlation function are thus encoded in the cross-correlation between \textit{standard} and relativistic contributions to the number counts. 

We now perform a multipole expansion in powers of the angle between the line of sight and the line joining the two pixels, denoted by $\beta_{ij}$. We perform the operation
\begin{equation}
\label{eq:multipoles}
    \xi_l(z,d)=\sum_{ij}\Delta(\mathbf{x}_i)\Delta(\mathbf{x}_j)\,P_l(\cos\beta_{ij}),
\end{equation}
where $P_l$ are the Legendre polynomials and the sum runs over the pair of pixels. We find that $\langle\Delta_{st}(\mathbf{x})\Delta_{st}(\mathbf{x}')\rangle$ only contribute to the \textit{even} multipoles $(l=0,\,2,\,4)$: we can combine the measurements of the monopole, quadrupole and hexadecapole to isolate $\delta$ and $\mathbf{V}$. This is what it has been done in current surveys \cite{boss}. The Doppler effects and the gravitational redshift contribute to the \textit{dipole} $(l=1)$ and \textit{octupole} $(l=3)$ alone, and therefore they break the symmetry of the correlation function \cite{gaz}. However, this contributions are only non-zero when we correlate two populations of galaxies with different luminosities. We have to split our population of galaxies into two families, the \textit{bright} and \textit{faint} galaxies \footnote{The choice is somewhat arbitrary, but the simplest case is to consider that half of our population has a luminosity above certain threshold (bright galaxies), while the other half has luminosity below the same luminosity threshold (faint galaxies).}. The other effects in $\Delta$, the ones of the third and fourth line in Eq.(\ref{eq:numbercounts}), are independent on the direction $\mathbf{n}$ and thus contribute only to the monopole of the correlation function. They are subdominant with respect to the density perturbations and can be neglected \cite{yoo,chall}. Therefore, by correlating two populations of galaxies with different luminosities, we can combine the even and odd multipoles of the correlation function to isolate the gravitational redshift contribution \cite{own,own2}.

\section{Measuring the anisotropic stress}

We have seen that the information about the distortions in the distribution of galaxies at large-scales is encoded in the multipoles of the correlation function. The different multipoles can be considered individual observables in its own right, since we can always fit the data to isolate them individually by weighting the correlation function with the correspondent Legendre polynomial (see Eq.(\ref{eq:multipoles})). The idea for the new test for the anisotropic stress is first to directly measure the dipole of the correlation function, and to combine it with the even multipoles to isolate $\Psi$. 

In practice, the dipole of the correlation function contains the terms $\langle\delta V\rangle$, $\langle V V\rangle$, $\langle\delta \dot V\rangle$ and $\langle\delta \Psi\rangle$. The $\langle\delta V\rangle$ and $\langle V V\rangle$ correlations are removed by using the even multipoles. The $\langle \delta \dot V\rangle$ correlation can be inferred from redshift evolution of the multipoles. This is the price to pay given that we do not want to assume the validity of the Euler equation. Putting all this together, we are left with the correlation between the density perturbations and the gravitational potential, i.e. the observable $O^{\delta\Psi}\equiv(b_B-b_F)\langle \delta \Psi\rangle$ \cite{own}.

Finally, measuring the anisotropic stress require an observable sensitive to $(\Phi+\Psi)$. This can be provided by \textit{galaxy-galaxy lensing}. Since $O^{\delta\Psi}$ is sensitive to the difference of the biases of bright and faint populations, we define a new observable that correlates the shapes of galaxies at high redshift (the lensed galaxies) with number counts of galaxies at low redshift (the lenses)
\begin{equation}
\label{eq:gglensing}
    O^{\delta(\Phi+\Psi)}\equiv\langle\Delta_B\kappa\rangle-\langle\Delta_F\kappa\rangle\propto(b_B-b_F)\langle\delta(\Phi+\Psi)\rangle,
\end{equation}
where $\kappa$ is the magnification of the lensed galaxies. We could also use the shear instead. Therefore, the ratio between this observable and the dipole measurement can give an estimator for the anisotropic stress $\eta=\Phi/\Psi$,
\begin{equation}
    O^\eta\equiv\frac{O^{\delta(\Phi+\Psi)}}{O^{\delta\Psi}}=\frac{(b_B-b_F)\langle\delta(\Phi+\Psi)\rangle}{(b_B-b_F)\langle \delta \Psi\rangle}\sim 1+\eta.
\end{equation}
Hence, the detection of any deviation from the value $2$ in the observable $O^\eta$ at any redshift will genuinely indicate the presence of an anisotropic stress. However, there is an important caveat here: we have implicitly assumed that the bias difference cancels. This means we need a survey which provides galaxy positions and shapes simultaneously, or the combination of two surveys that probe the same catalog of galaxies. While this is unlikely for the current and future  (Euclid or SKA), we have found that there is no significant statistical difference if we measure $\langle \delta \Psi\rangle$ alone, which removes the need to define $O^{\delta(\Phi+\Psi)}$ as in Eq.(\ref{eq:gglensing}) \cite{own2}.

\section{Conclusions \& Outlook}

Next generation of galaxy surveys will provide us with an enormous amount of data which can be used to test the laws of gravity on scales that are currently inaccesible. This opens the door to construct new observables which probe effects in the large-scale distribution of galaxies that are neglected in current analysis. On top of the density perturbations, redshift-space distortions and weak lensing, we have shown that the galaxy number counts at linear order in perturbation theory is also affected by relativistic distortions. 

The relevant relativistic effects consist in two Doppler contributions and the gravitational redshift. They break the symmetry of the correlation function and contribute to the odd multipoles. In order to detect them, we must correlate galaxies with different luminosities. Given that each of the multipoles can be considered an individual observable, it is possible to combine even and odd multipoles to isolate the relativistic effects. This provides a direct measurement of $\langle\delta\Psi\rangle$, for which we expect an accuracy of $20-30\%$ in SKA at low redshift and large separations between galaxies \cite{own2}. Finally, the combination of this measurement with galaxy-galaxy lensing observations, which are sensitive to $\langle\delta(\Phi+\Psi)\rangle$, can be used to build an estimator for $\eta$ \cite{own3}. This method is model independent, in the sense that it does not assume the validity of the Euler or continuity equations, and thus provides a genuine measurement of the anisotropic stress.

\section*{Acknowledgments}

This project and the presence of the author at the conference has received funding from the European Research Council (ERC) under the European Union's Horizon 2020 research and innovation program (Grant agreement No. 863929; project title ``Testing the law of gravity with novel large-scale structure observables"). We also acknowledge funding from the Swiss National Science Fundation (SNSF). The author also acknowledge the organization of the ``56th Rencontres de Moriond" conference for the opportunity to present this work.


\section*{References}

\begin{thebibliography}{99}

\bibitem{amen} L. Amendola \emph{et a.l.}, \Journal{\PRD}{87}{023501}{2013}.

\bibitem{own} D. Sobral Blanco and C. Bonvin, \Journal{\PRD}{104}{063516}{2021}

\bibitem{bon} C. Bonvin and R. Durrer, \Journal{\PRD}{84}{063505}{2011}.

\bibitem{yoo} J. Yoo \emph{et a.l.}, \Journal{\PRD}{80}{083514}{2010}.

\bibitem{chall} A. Challinor and A. Lewis \Journal{\PRD}{84}{043516}{2011}.

\bibitem{kai} N. Kaiser, \Journal{\emph{Mon. Not. Roy. Astron. Soc.}}{227}{1}{1984}.

\bibitem{ham} A. J. S. Hamilton in \emph{Ringberg Workshop on Large Scale Structure, Ringberg, Germany, September 23-28, 1996}, (1997).

\bibitem{boss} BOSS: Sapathy \textit{et a.l.}, \Journal{\emph{Mon. Not. Roy. Astron. Soc.}}{469}{2}{2017}.

\bibitem{gaz} C. Bonvin, L. Hui and E. Gaztañaga, \Journal{\PRD}{89}{083535}{2014}.

\bibitem{own2} D. Sobral Blanco and C. Bonvin, \emph{in preparation}.

\bibitem{own3} I. Tutusaus, D. Sobral Blanco and C. Bonvin \emph{in preparation}.

\end{thebibliography}
\end{document}